\documentclass[aps,superscriptaddress,showpacs,preprint]{revtex4}
\usepackage[dvips]{graphicx}

\begin{document}

\title{Berry-Phase Induced Dynamical Instability and Minimum Conductivity in Graphene}

\author{Shi-Jie Xiong}
\email{sjxiong@nju.edu.cn}
\affiliation{National Laboratory of
Solid State Microstructures and Department of Physics, Nanjing
University, Nanjing 210093, China}
\author{Ye Xiong}
\affiliation{College of Physical Science and Technology, Nanjing
Normal University, Nanjing 210097, China}

\begin{abstract}

Single-layer carbon, or graphene, demonstrates amazing transport
properties, such as the minimum conductivity near $\frac{4e^2}{h}$
independent of shapes and mobility of samples. This indicates there
exist some unusual effects due to specific Dirac dispersion relation
of fermion in two dimensions. By deriving fermion-lattice
interaction Hamiltonian we show that Berry phases can be produced in
fermion states around two Dirac points by relative rotations of two
sublattices. The Berry phases in turn remove the degeneracies of
energies for states near the Fermi surface, leading to a dynamical
instability of the lattice with respect to the rotations. By
considering the Berry phases emerging in an uncertain way on fermion
wavefunctions in vicinities of the Fermi surface, the conductivity
is calculated by using the Landauer-B\"{u}tticker formula together
with the transfer-matrix technique, verifying $\sim \frac{4e^2}{h}$
quantized minimum conductivity as observed in experiments
independent of shapes and sizes. The relationship between the
chaotic structure of fermions due to the Berry phases and the
classical transport properties are discussed. The physical meaning
is profound as this relationship provides an excellent example to
elucidate the mechanism of quantum-classical transition.

\end{abstract}


\maketitle


As a potential candidate material for carbon-based electronics, the
graphene, a two-dimensional (2D) honeycomb lattice of carbon atoms,
has attracted much attention \cite{1,2,3,4,5}. It is found that both
the type (electrons or holes) and the number of carriers can be
tuned by changing the gate voltage \cite{3,7}. Such tunability
originates from the specific dispersion relation of the massless
Dirac fermions in graphene \cite{8a,8}. The Fermi level of the
system can be varied by changing the gate voltage. Particularly,
there exists a universal maximal resistivity of the order of
6.5k$\Omega$ for all samples with the Fermi level near the Dirac
points, independent of their mobility. By increasing the Fermi level
the conductivity linearly increases \cite{3}. The value of the
minimal resistivity ($\frac{h}{4e^2} \sim 6.5\text{k}\Omega$)
\cite{3} implies the existence of at least 4 channels (including
spin degrees of freedom) in a square of unit area and the number of
channels increases with increasing the sample width. Although there
are 4 ballistic channels at a node of the Dirac dispersion relation
in a graphene lattice, the experimental fact of constant
conductivity is inconsistent with the quantum-mechanical description
from which all states are localized by extremely weak disorder in 2D
\cite{9}. The non-zero minimal conductivity has been investigated by
several theoretical studies using different methods
\cite{a6,a8,a9,a10}. The minimum conductivity was correctly
demonstrated, but as to the reason for it and as to its exact value
being $e^2/h$ or $e^2/\pi h$ there is no consensus at the present.
By carefully examining the interaction between Dirac fermions and
lattice motions, we show that the Berry phases produced by relative
rotations of two sublattices can induce dynamical instability and
cause constant conductivity $\frac{4e^2}{h}$.

The main features of the band structure in graphene can be well
described by a tight-binding Hamiltonian with one $\pi$ orbital per
site on the graphene lattice \cite{8a}:
\begin{equation}
  H= \sum_{\langle n,n' \rangle} t_{nn'} (a^\dag_n a_{n'} +a^\dag_{n'} a_n),
  \end{equation}
where $a^\dag_n$ ($a_n$) creates (annihilates) an electron at site
$n$, $\langle \ldots \rangle$ denotes the nearest-neighbor (NN)
sites, and $t_{nn'}$ is the NN hopping. Here, we omit the diagonal
terms by shifting the energy zero as no diagonal disorder is
considered, and the spin indices are not explicitly included. The
hopping integral $t_{nn'}$ depends on distance between NN sites.
Considering a relative displacement ${\bf d}\equiv (d_x,d_y)$
between two sublattices as shown in Fig. 1, the lengths of three
bonds in a cell become
\begin{equation}
   l_i=\sqrt{[d \sin (\theta-\alpha_i)]^2+[l_0-d \cos (\theta-\alpha_i)]^2}, \,\,
\end{equation}
where $\theta$ is the angle of ${\bf d}$ related to $x$ axis, $l_0$
is the original bond length, and $\alpha_i = 0, \frac{2\pi}{3},
-\frac{2\pi}{3}$ for $i=1,2,3$. For small displacement, we only keep
the first-order terms of $d$ and obtain $l_i \sim l_0-d \cos
(\theta-\alpha_i)$. Then the corresponding hopping integrals can be
written as
\begin{equation}
   t_i =t_0 +\lambda d \cos
(\theta-\alpha_i),
\end{equation}
where $t_0$ is the original hopping and $\lambda$ is a coefficient
describing the linear dependence of $t_i$ on $l_i$ in the case of
small $d$. On the other hand, the elastic energy trends to keep
$d=0$. The in-plane displacements occur in two vibrational modes
along $x$ and $y$ directions. We adopt variables $d_x$ and $d_y$ to
express these two vibration modes. Using the Bloch transformation
for graphene lattice and keeping terms of the first orders of $d_x$,
$d_y$, and electron wave vector ${\bf k}$ related to the Dirac
points, the Hamiltonian becomes
\begin{equation}
 \label{ham}
   H= \frac{3}{2} \sum_{\bf k} \left[ (t_0 k_y \hat{\tau}_z +
   \lambda d_x \hat{\textbf{1}}) \otimes \hat{\sigma}_x -(t_0 k_x \hat{\textbf{1}}
   -\lambda d_y \hat{\tau_z}) \otimes \hat{\sigma}_y \right],
   \end{equation}
where $\hat{\textbf{1}}$ and $\hat{\tau}_z$ are unit and Pauli
matrices acting on two irreducible Dirac points, and
$\hat{\sigma}_{x,y}$ are Pauli matrices on two sublattices. So the
dispersion relations around two Dirac points are
\begin{equation}
  \label{ener}
   E^{(\pm)}_{1} =\pm \frac{3}{2} \left| t_0 (k_y+\text{i}k_x) +\lambda
   ( d_x-\text{i} d_y) \right|, \, E^{(\pm)}_{2} =\pm \frac{3}{2}
   \left| t_0 (-k_y+\text{i}k_x) +\lambda
   ( d_x+\text{i} d_y) \right|.
   \end{equation}
\begin{figure}[ht]
\includegraphics[width=12.6cm]{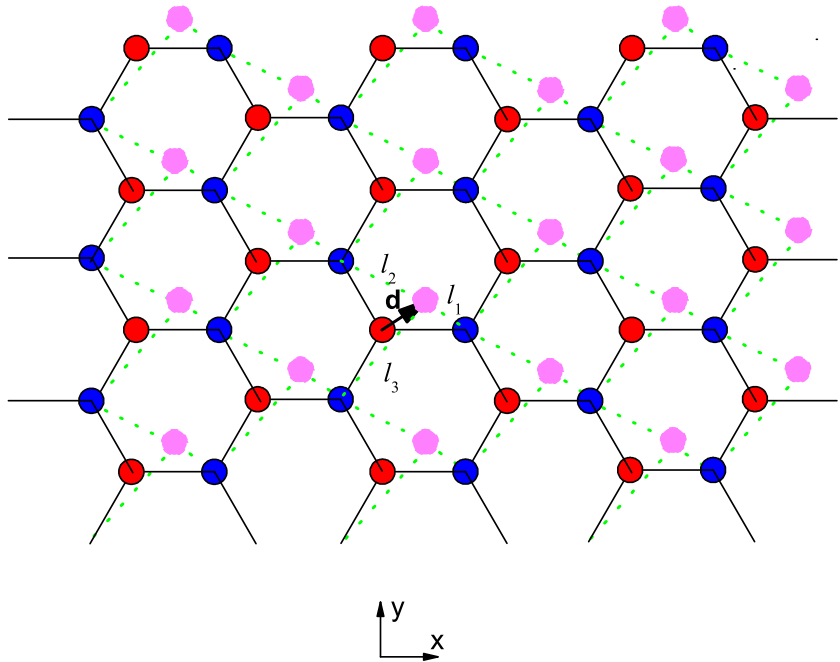}
\caption {Sublattices A (red points) and B (blue points) on
graphene lattice. With relative displacement ${\bf d}$, sublattice
A is moved to positions marked with light magenta circles. }
\label{fig1}
\end{figure}

It is interesting to note that the diabolical points of the
dispersion relations are moved by sublattice displacements. The
dependence of energies on real ${\bf d}$ for a given ${\bf k}$ also
has a diabolical point determined by $(d_x=-t_0k_y/\lambda,d_y=t_0
k_x/\lambda)$ for $E_{1}^{(\pm)}$ and by
$(d_x=t_0k_y/\lambda,d_y=-t_0 k_x/\lambda)$ for $E_{2}^{(\pm)}$ as
shown in Fig. 2.

\begin{figure}[ht]
\includegraphics[width=12.6cm]{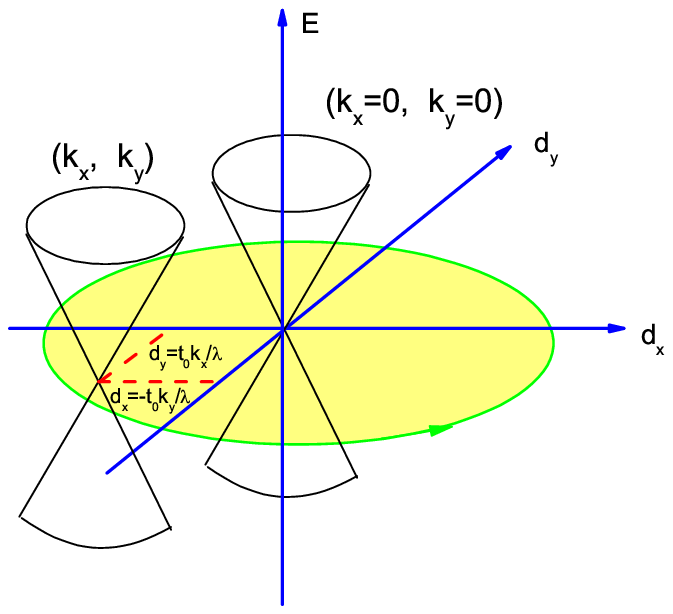}
\caption {$E^{(\pm)}_{1}$ as functions of ${\bf d}$ for different
values of ${\bf k}$. } \label{fig2}
\end{figure}

From the Berry theorem a circle of rotation in a parameter space can
cause a Berry phase $\pm \pi$ in a wavefunction whose eigenenergy
has a diabolical point enclosed in this circle \cite{berry}. If
there exists a circular vibration with amplitude $d_f =t_0 k_f
/\lambda$ with $k_f$ being the magnitude of the Fermi wave vector,
in every circle of the vibration the states inside the Fermi circle
change sign due to the acquired Berry phases but the states outside
the Fermi circle do not. Now we consider an inside state
$|\psi_1\rangle$ and an outside state $|\psi_2 \rangle$ situated in
an infinitesimal vicinity of a ${\bf k}$ point at the Fermi circle.
They can be degenerate because of the continuity of the band. So we
can construct a $\psi_+$-$\psi_-$ basis to represent these two
states:
\begin{equation}
   |\psi_\pm \rangle =\frac{1}{\sqrt{2}} ( |\psi_1\rangle \pm
   |\psi_2\rangle  ).
\end{equation}
The asymmetric acquirement of a $\pm\pi$ Berry phase in
$|\psi_{1,2}\rangle $ during a circle of circular vibration $d_f$
leads to a hybridization of $\psi_\pm({\bf k})$ and removes the
degeneracy. This can be described by an effective Hamiltonian
\begin{equation}
   H_B =  \sum_{\alpha =\pm} |\psi_\alpha
   \rangle \epsilon_f
   \langle \psi_\alpha | +
   g (|\psi_+  \rangle \langle \psi_- | +\text{H.c.}) ,
\end{equation}
where $\epsilon_f $ is the Fermi energy and $g$ is the hybridization
strength depending on the rotation rates. Hamiltonian $H_B$ raises
one level to $\epsilon_f+g$ and lowers the other to $\epsilon_f-g$.
Since the former is empty and the latter is occupied by fermion,
this leads to the lowering of the fermion energy by $g$. Thus, the
energy gain per cell of the fermion system due to the vibration
$d_f$ is $E_g \sim 2\pi k_f g D$ where $D$ is the width of a narrow
stripe around the Fermi circle in which the fermion energies are
affected by $g$. At the same time the energy per cell of the
circular vibration $d_f$ is proportional to $d_f^2$ and can be
written as $E_v\sim c d_0^2$ with $c$ being a prefactor. So, if $E_v
\leq E_g$, viz.
 \begin{equation}
 k_f \leq k_v \equiv \frac{2\pi g D \lambda^2}{ct_0^2},
 \label{con}
 \end{equation}
a dynamical instability occurs with undamped fluctuations of
rotations and Berry phases.

The condition (\ref{con}) for the dynamical instability can be
easily satisfied since $k_v$ is finite and positive while $k_f
\rightarrow 0$ for the Fermi level situated at the Dirac points. It
could still be satisfied even by slightly shifting the Fermi level
with the gate voltage $V_g$. The value of $k_v$ can be estimated as
follows: From the Harrison formula, the hopping integral of NN sites
$n$ and $n'$ is $t_{nn'} =t_0 \left( \frac{l_0}{l_{nn'} }\right)^2$
\cite{harri}. So $\lambda \sim \frac{2t_0}{l_0} \sim 1.4 t_0
$\AA$^{-1}$. $c$ is related to the elastic coefficient $s_e$ of the
C-C bonds by $ c\sim \frac{3}{4} s_e$, and $s_e \sim 43.37$eV
\AA$^{-2}$ \cite{se}. $g$ is related to the vibration frequency
$\omega_0$ by $g\sim \hbar \omega_0$, and $D$ is mainly due to the
uncertainty of the Fermi energy in the vibration and can be
expressed as $W\sim \frac{2\hbar \omega_0}{3t_0}$. From the value of
$s_e$ and the mass of carbon atom we have $\hbar \omega_0 \sim
0.1$eV. All these can give $k_v\sim 10^{-3}$ in units of $l_0^{-1}$.
This value is considerable as the whole Dirac dispersion relation is
valid for small $k$.

The circular vibrations could also propagate in the space with
different wavelengths longer than the lattice spacing
(long-wavelength limit) and cause mesoscopic inhomogeneity of the
created Berry phases. The spatial and temporal uncertainty of Berry
phases attached on fermion states by such circular vibrations has
radical effects on the basic structure of fermions in graphene. To
illustrate this we adopt a model with random signs of NN hoppings
$t_{nn'} =t_0$ or $-t_0$ to mimic the fluctuations of Berry phases
$\pm \pi$ and calculate the conductivity for a rectangular sheet
with different width $W$ and length $L$ connecting to two reservoirs
with Fermi energy at the Dirac points ($E_f=0$) by using the
Landauer-B\"{u}ttiker formula:
\begin{equation}
 \label{sigma}
   \sigma = \frac{2 L e^2}{W h} \sum_l \frac{1}{\cosh^2 ( \gamma_l
   L)} ,
   \end{equation}
where $\gamma_l$ is the Lyapunov exponent of the $l$th channel in
the system with width $W$ calculated with the transfer-matrix
technique. The prefactor $2$ is from contributions of two spins. The
obtained conductivity as a function of $W/L$ for different sizes is
shown in Fig. 3. Except for small values of $W/L$, the conductivity
has almost a constant value near $\frac{4e^2}{h}$ independent of
$W/L$ and $W$, in consistence with the experimental findings.
Surprisingly, this value is even larger than $\frac{4e^2}{\pi h}$
obtained in ballistic graphene \cite{perf} in spite of the disorder
introduced by random signs of hoppings. In a perfect 2D Dirac
fermion gas, the difference between two successive Lyapunov
exponents in Eq. (\ref{sigma}) is $\delta \gamma = 2\pi /W$ due to
the Dirac fermion dispersion relation \cite{perf}. In the present
model the values of $W \delta \gamma$ are distributed within
$[1,2.5]$, smaller than that in the perfect case by a factor about
$\pi$, leading to the $\pi$ times of the conductivity as can be
derived from the expansion of Eq. (\ref{sigma}). Thus, the $\pi$
factor difference in the conductivity is not trivial and reflects an
essential change of the fermion properties due to the Berry-phase
induced instability.

\begin{figure}
\includegraphics[width=12.6cm]{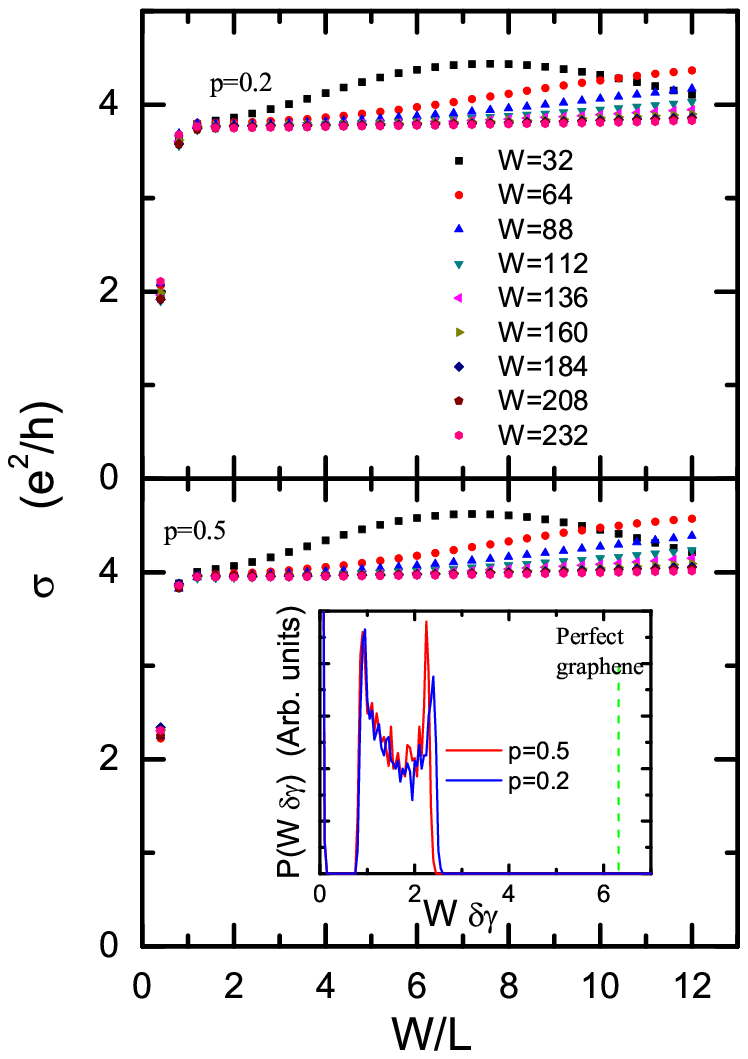}
\caption{Calculated conductivity as a function of ratio $W/L$ for
rectangular sheets with different transverse sizes $W$. Inset in the
lower panel: Distribution function $P( W\delta \gamma)$ of $W\delta
\gamma$. The green dashed line indicates the $\delta$-function
distribution at $2\pi$ in the case of perfect sheet. Units of $W$
and $L$ are cell sizes of graphene lattice. } \label{fig3}
\end{figure}

Besides the phase fluctuation, there are complicated level crossings
in the dependence of energy on ${\bf k}$ and ${\bf d}$. The energy
degeneracies at crossings can be easily modified by slight
disturbing, resulting in chaotic energy structure. Thus, the global
quantum-mechanical description of the Dirac fermions could be
modified with a semi-classical scenario: The whole sheet can be
divided into areas of size $b_0$ which are classically connected
together as there is no quantum coherence among them. This is
equivalent to an ohmic network of small conductors each of which
represents an area of size $b_0$. Thus, the conductivity of the
whole system is determined by the conductivity of every area which
can be quantized. For the Fermi energy $E_f=0$ there are only two
transmission channels in an area providing minimum conductivity
$\sigma = \frac{4e^2}{h}$. By applying the gate voltage $V_g$ the
Fermi energy is shifted as $E_f = \beta V_g$ with $\beta$ being a
prefactor and the number of transmission channels in an area is
$\frac{2\beta |V_g| b_0}{3\pi t_0}$, producing a conductivity
$\sigma\sim  \frac{4e^2 \beta b_0 |V_g| }{3\pi h t_0}$. The
experiments show a linear dependence of $\sigma $ on the gate
voltage $V_g$ with a finite slope, suggesting a finite coherent size
$b_0$. From the slope one could estimate $\frac{\beta b_0 }{3\pi
t_0} \sim 0.2 V^{-1}$.

In summary, we show the specific coupling between the Dirac fermions
and the circular vibrations in graphene which can lead to a
dynamical instability due to the produced Berry phases. The
calculations from this scenario provide a constant conductivity
consistent with the experiments. The results indicate that a picture
of an ohmic network consisting of small quantum conductors is
suitable to account for the peculiar transport properties in
graphene.

{\it Acknowledgments} This work was supported by the State Key
Programs for Basic Research of China (2005CB623605 and
2006CB921803), and by National Foundation of Natural Science in
China Grant Nos. 10474033 and 60676056.


\end{document}